# iDriveSense: Dynamic Route Planning Involving Roads Quality Information

Amr S. El-Wakeel*, Aboelmagd Noureldin*†‡, Hossam S. Hassanein*‡ and Nizar Zorba†

*Electrical and Computer Eng. Dept., Queen's University, Kingston, ON, Canada, K7L 3N6
†Electrical and Computer Eng. Dept., Royal Military College of Canada, Kingston, ON, Canada, K7K 7B4
‡School of Computing, Queen's University, Kingston, ON, Canada, K7L 3N6
†Electrical Eng. Dept., Qatar University, Doha, Qatar, P.O. Box 2713
Email: {amr.elwakeel, nourelda}@queensu.ca, hossam@cs.queensu.ca, nizarz@qu.edu.qa

*Abstract*—Owing to the expeditious growth in the information and communication technologies, smart cities have raised the expectations in terms of efficient functioning and management. One key aspect of residents' daily comfort is assured through affording reliable traffic management and route planning. Comprehensively, the majority of the present trip planning applications and service providers are enabling their trip planning recommendations relying on shortest paths and/or fastest routes. However, such suggestions may discount drivers' preferences with respect to safe and less disturbing trips. Road anomalies such as cracks, potholes, and manholes induce risky driving scenarios and can lead to vehicles damages and costly repairs. Accordingly, in this paper, we propose a crowdsensing based dynamic route planning system. Leveraging both the vehicle motion sensors and the inertial sensors within the smart devices, road surface types and anomalies have been detected and categorized. In addition, the monitored events are geo-referenced utilizing GPS receivers on both vehicles and smart devices. Consequently, road segments assessments are conducted using fuzzy system models based on aspects such as the number of anomalies and their severity levels in each road segment. Afterward, another fuzzy model is adopted to recommend the best trip routes based on the road segments quality in each potential route. Extensive road experiments are held to build and show the potential of the proposed system.

## I. INTRODUCTION

Smart Cities, by 2024, are predicted to generate $2.3 trillion according to CISCO [1]. Meanwhile, there are various smart applications and services present in multiple sectors spanning environment, health, waste management and transportation [2, 3]. Nevertheless, further insights, evaluations, and improvements are necessary for granting adequate performance of smart cities. Mainly, smart transportation and traffic management are highly needed as one can say they broadly influence almost every aspect of the smart cities operation on daily bases [4]. In particular, trip route planning receives great interest particularly in big and crowded cities [4, 5]. Principally, trip planning applications and service providers afford route recommendations based on relatively shorter paths, traffic congestion and even with up to date construction works [6].

Consequently, some of the route planning key players as Google have adopted online dynamic routing driven by live traffic network information. For example, Google maps provide an online suggestion for vehicle re-routing when roads are experiencing instantaneous traffic congestion based on many factors such as untraditional mobility behavior or accidents. On the other hand, corwdsensed based trip planning application Waze [7] relies on lively sensed traffic situations which shared with the users' intervention. APOLO [8] system was introduced to overcome the network overload introduced in many of the traffic management systems because of information exchange between vehicles and servers. This system proposed a centralized traffic monitoring system that works on both online and offline bases. In the offline stage, mobility patterns are conducted by historical data processing while in the online stage vehicles are re-routed away from the congested routes. The results showed travel time reduction of 17 % along with a speed increase of 6% compared to present approaches.

In addition, various efforts in the literature provided suggestions to enable shorter and faster routing for land vehicles. In [9], an adaptive routing approach was introduced and dealt with route planning as a probabilistic dynamic problem. In their algorithms, they aimed to reduce the predicted en route trip time while considering broadcasted traffic information, onboard based traffic state measurement, and historical traffic patterns. Moreover, in [10] personal behavior based trip planning was presented to contribute a solution for traffic congestion problem. The authors assumed and discussed that the driving preferences changes from a driver to another could be handled in a way to create drivers' profiles which are used in their route planning leading to less traffic congestion. Furthermore, in [6] an extended version of [10] specified three significant aspects of personal based route planning. These significant considerations are the road safety regarding the presence of snow or black ice, traffic speed and congestion level. The contributions of these factors are assessed based on fuzzy inference engine while the overall optimum routing was enabled by an optimization problem. In [11] a dynamic route planning system was proposed to include future traffic hazards in vehicle routing. This system contained three components which are real-time data streamed from the vehicles plus data collected by automatic traffic loops

This research is supported by a grant from the Natural Sciences and Engineering Research Council of Canada (NSERC) under grant number: STPGP 479248. In addition, this work was made possible by NPRP grant NPRP 9-185-2-096 from the Qatar National Research Fund (a member of The Qatar Foundation).

sensors and the third component used both data sources to predict future traffic conditions through Spatio-Temporal random field process.

In order to assure relaxing trips, in [12] a system was introduced to reduce the routes distances along with providing suggestions for routes with high-quality sceneries. A memetic algorithm was used to provide skyline scenic trip planning while maintaining low travel distances. On the other hand, to ensure drivers and travelers safety, a system was proposed in [13] to enable route planning while discarding routes that encounter high crime rate. Based on crime data provided by Chicago and Philadelphia a risk model was introduced for the cities urban networks.

The highlighted literature showed significant efforts in enriching efficient route planning regarding suggesting shortest paths, fewer traffic routes and considering personal preferences as well. However, road quality information is the crucial aspect that enables drivers safe and comfort trips was not considered in the most of route planning systems [14]. Deteriorated road surface conditions can lead to vehicle damage and dangerous driving scenarios that result in drivers' frustration and stress [14, 15]. Consequently, existing land vehicles are considered mobile sensor hubs with various sensing and communications capabilities [16]. Vehicle motion sensors in the land vehicles along with the inertial sensors embedded in the drivers' smart devices enabled adequate detection for various road surface types and anomalies. Thanks to both GPS receivers and inertial sensors the detected anomalies are robustly geo-referenced [17, 18].

In this paper, we present iDriveSense a corwdsensed based Road Information Services (RIS) system. In this system, we leverage the sensing capabilities of the land vehicles and drivers' smart devices to generate detailed data sets of the road surface types and anomalies with different severity levels. Also, these datasets are used as an input to a cloud-based Fuzzy Inference System (FIS) utilized for road segment assessments. Moreover, iDriveSense provides independent route planning or through evaluating potential routes suggested by trip planning service providers like Google Maps. Route suggestions and evaluations are enabled to the drivers through another FIS.

## II. SYSTEM STRUCTURE

In this section, we present the system configuration used to build iDriveSense. As mentioned earlier, in this system land vehicles are considered as mobile crowdsensing nodes. As shown in fig. 1, detailed and descriptive datasets of road surface conditions are sent to a cloud RIS. Accordingly, road segments assessment and route recommendations are provided through cascaded FIS.

### A. FIS

Basically, fuzzy logic is intended to deal with real-world applications through framework able to deal with ambiguity and inaccuracy [19]. In fuzzy logic, quantified rules or statements are adopted to avoid firm true or false decisions. Accordingly, fuzzy logic sets grant objects values that range from 0 to 1 through graded memberships. Therefore, FIS maps sets of given inputs to outputs with the aid of fuzzy logic.

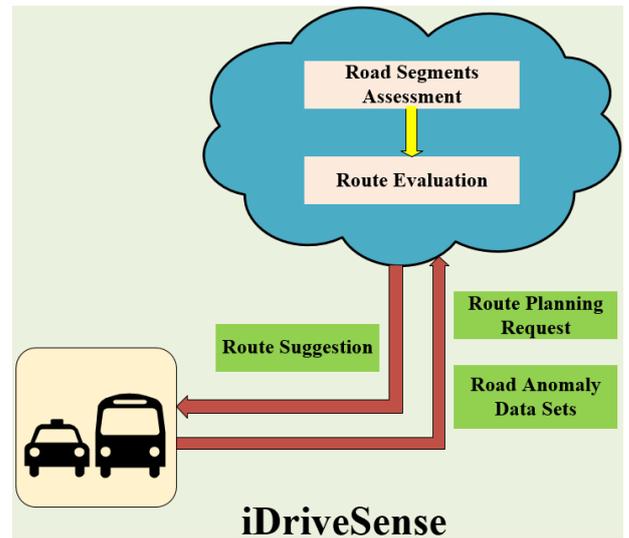

Fig. 1 iDriveSense system architecture

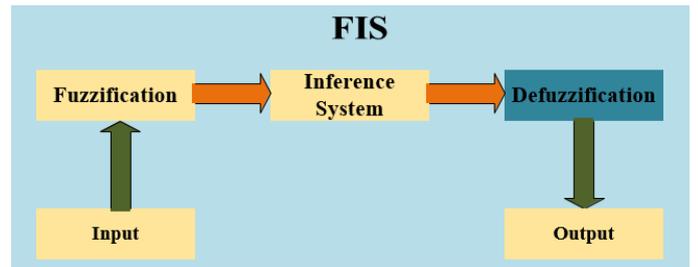

Fig. 2 FIS system Structure.

In addition, FIS dynamic performance is modeled by sets of linguistic descriptive rules that are set according to the system designer prior knowledge [19]. For example, the fuzzy rules of a multiple-input-single-output (MISO) fuzzy system are given by

$\textbf{R1}$: if ($a$) is X1 and ($b$) is Y1, then ($c$) is Z1;

$\textbf{R2}$: if ($a$) is X2 and ($b$) is Y2, then ($c$) is Z2;     (1)

……….

$\textbf{Rn:}$ if ($a$) is Xn and ($b$) is Yn, then ($c$) is Zn;

As $a, b$ and $c$ are linguistic variables representing two inputs process state variables and one output variable. While, Xi and Yi are linguistic values of the linguistic $a, b$ in the universe of discourse U and V with $i = 1, 2, …, n$. The linguistics values Zi of the linguistic variable $c$ in the universe of discourse $W$ in case of Mamdani FIS [20].

Fundamentally, as shown in fig. 2, four components together represent the FIS. The fuzzy rules which can be called "IF-THEN" are built according to the prior knowledge of the required system. Also, the input domain crisp values U are outlined with fuzzy sets defined in the same universe of disclosure by the aid the fuzzification stage. On the other hand, an inverted operation is carried by the defuzzification stage to map the crisp values of the output domain V with the predefined fuzzy sets. Further details on FIS structure and derivations can be found in [21].

## B. Road Segment Assesment FIS

For road segment assessments, inputs from the road surface types and conditions data sets are used to compute three inputs for the FIS. The first input is the related to the total number of road anomalies in a given road segment $S$. For each segment, a normalized percentage of road anomalies $RA$ is computed simple through dividing the total number of anomalies over $S$ to reflect the density of the anomalies in particular segment. Thus this input is mapped to three membership functions which are defined as low, moderate and high. As shown in fig.3 we adopted *sigmoidal* membership function for both low and high functions. A *sigmoidal* function is a mapping on input vector $a$, and can be represented by:

$$f(a,m,n) = \frac{1}{1+\exp(-m(a-n))} \quad (2)$$

Where the *sigmoidal* membership functions innately open to the right or left according to the sign of the parameter $m$ and $n$ is a control parameter. The product of two *sigmoidal* functions is used in the moderate function and is given by:

$$f_k(a) = \frac{1}{1+\exp(-m_k(a-n_k))} \quad (3)$$

Given that $k = 1,2$ and the parameters $m_1$ and $m_2$ command the left and right curves slopes and these two parameters have to be positive and negative, respectively. While $n_1$ and $n_2$ control the left and right curves points of inflection.

The second input is representing the effect of the anomalies severity level on the assessment of a road segment. As the road segments with equal lengths and have the same density of anomalies should not receive the same assessment decision if they experience different types of anomalies with different levels of severity. Accordingly, the average percentage of anomalies severity level in each segment is calculated and normalized concerning the segment length and presented by mild and severe *sigmoid* membership functions. Lastly, the third input is to distinguish road segments of single and double lanes. This input was chosen to represent the significance of the road segment wideness on its quality assessment. The road segments with multi-lanes allow the driver to maneuver before the anomalies easily while this is difficult to occur in single road segments and it can lead to dangerous scenarios within the two ways road segments. The third input is also mapped through two *sigmoid* wide and narrow membership functions. In this FIS, the fuzzification of the inputs is mapped by 11 Mamdani fuzzy rules. The road assessment FIS is then defuzzified to enable three output levels of road segment quality. They are classified into Good, Moderate and poor segments.

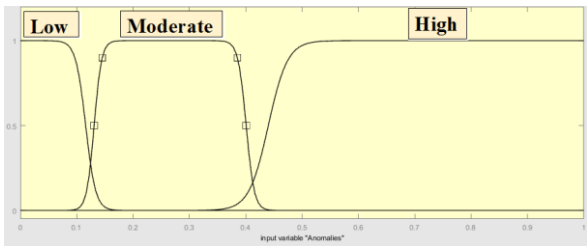
(a)

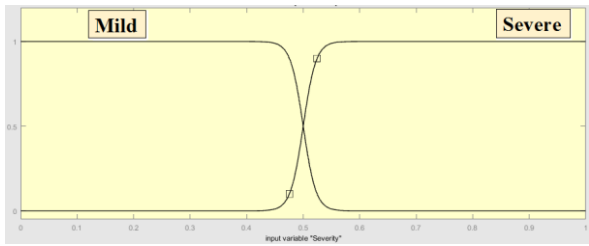
(b)

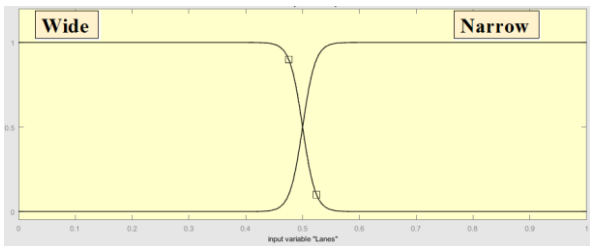
(c)

Fig.3 Membership functions utilized in road segments assessment: a) percentage of anomalies (low, moderate, and high), b) average level of severity (low and high) and c) lanes (narrow and wide)

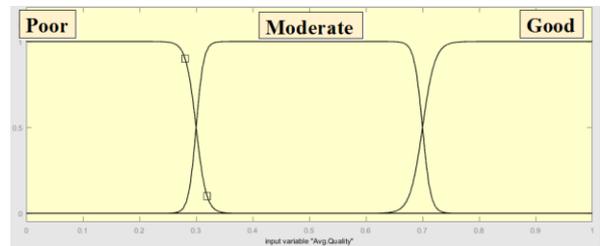
(a)

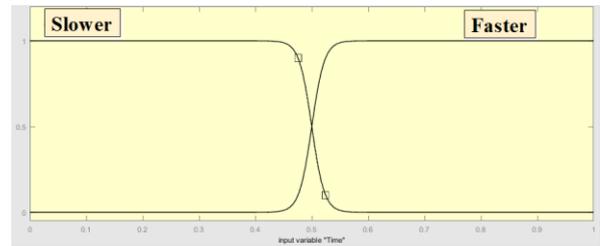
(b)

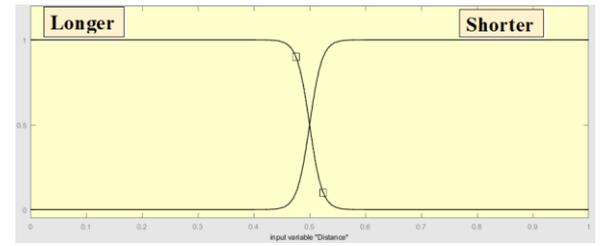
(c)

Fig.4 Membership functions utilized in route suggestion: a) average segments quality (poor, moderate, and good), b) route time (slow and fast) and c) route length (long and short)

## C. Route Suggestion FIS

Regarding the route suggestions, a cascaded FIS is utilized as shown in fig. 1 to provide route recommendation and evaluation. In this FIS, as shown in fig. 4, there are three inputs adopted to decide the route recommendation. The first input is the average quality of the segments in a potential route. This input is controlled by three membership functions namely poor, moderate and good which reflects one aspect of the route evaluation. We adopted two *sigmoid* functions for the poor and good membership functions while we used the product of two *sigmoidal* functions for the moderate one. It is worth mentioning that the primary concern in iDrivesense route planning is in providing high road quality routes. However, high traffic routes and long paths should be avoided as well whenever is possible. Therefore, the second and the third inputs are described by the route travel time and route distance, respectively. The second one is divided by two sigmoid membership functions named slower and faster. On the other hand, the third input is also described by two sigmoid membership functions called longer and shorter. In the route recommendation FIS, the fuzzification of the three inputs is controlled by 12 fuzzy rules. While the defuzzification of this cascaded FIS provides three output levels of route recommendations. They are divided into (not suggested, marginally suggested and suggested).

## III. RESULTS AND DISCUSSION

In order to assess the performance of the proposed system, we conducted extensive experiments in Kingston, ON, Canada. These experiments were held adopting multiple vehicles and included various motion sensors and smart devices. These road experiments involved numerous roads in heavy traffic downtown core, urban and neighborhoods residential areas to assure the variety of road segments quality and routing approaches. Accordingly, to show the performance of the iDriveSense system in route recommendation considering the road quality information. We consider a real trip request as shown in fig. 5. In this trip, the driver requires route planning to travel from point A to point B while requesting a stable and safe drive as the highest priority. According to Google maps, as shown in fig. 6, there are two recommended routes. The first one reaches point B in 5 minutes, and it is 1.3 Km regarding route distance. On the other hand, the second recommended route travel time is 7 minutes with a distance of 1.4 Km. Thus according to Google maps suggestions which are mainly provided based on less trip time and shortest route distances, Route 1 is recommended as shown figure.6.

Consequently, as requested by the driver, the safe and high road quality has the highest priority in the trip satisfaction. For this regard, iDriveSense examined the quality of the road segments in the potential routes as listed Table 1 and shown in fig. 7. With the aid of the first FIS system described in Sec. II., the road segments quality of Route 1 and Route 2 were assessed. The first route (suggested by Google Maps) has nine road segments.

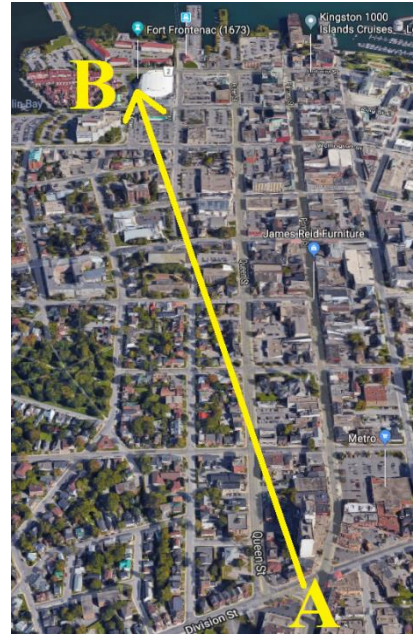

Fig.5 Route planning request from point A to point B (Top view).

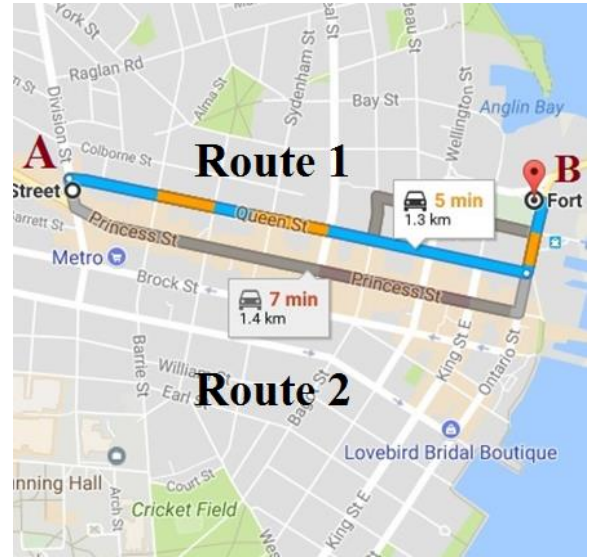

Fig 6. Route suggestions provided by Google Maps.

**Table I. Road Segment Assessment for Route 1 and Route 2**

| ROAD SEGMENTS | ROUTE 1 | ROUTE 2 |
|---|---|---|
| 1 | Poor | Moderate |
| 2 | Poor | Moderate |
| 3 | Poor | Good |
| 4 | Poor | Good |
| 5 | Poor | Good |
| 6 | Poor | Good |
| 7 | Poor | Good |
| 8 | Moderate | Moderate |
| 9 | Poor | Moderate |
| 10 | NA | Poor |

In this route, there are eight segments assessed as poor quality ones, and there is only one segment assessed as a moderate one. On the other hand, the second recommended route (Route 2) consists of 10 segments. Utilizing the first FIS, 5 of the road segments within this route are evaluated as good ones, and there were other four assessed as moderate while only one is considered a poor road segment. As per fig. 7, the assessed road segments of Route 1 and Route 2 are highlighted with different colors to indicate different levels of quality.

Afterward, the route suggestion FIS was adopted to recommend the route with high road segments quality. In this cascaded FIS, the output of the first FIS along with the trip time and route distance in each route is used to set the routes recommendation levels. Given the predefined fuzzy rules, the three inputs and the required priority to the route with high road segments quality, iDrivesense contrary to Google Maps has recommended Route 2 as shown in fig. 8

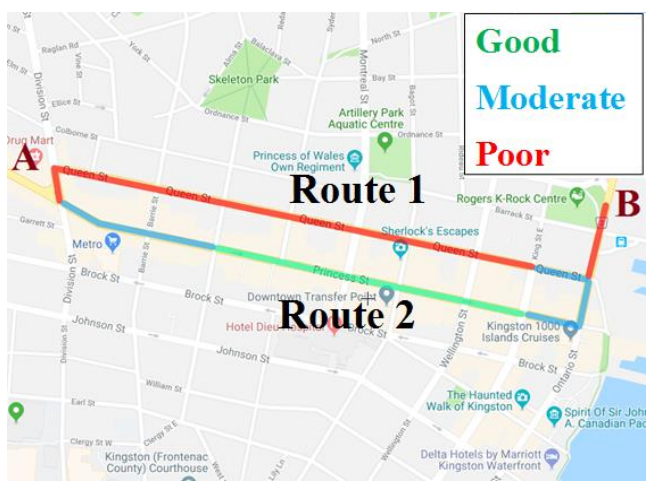

Fig 7. Road segments assessment by iDriveSense for the routes suggested by Google Maps.

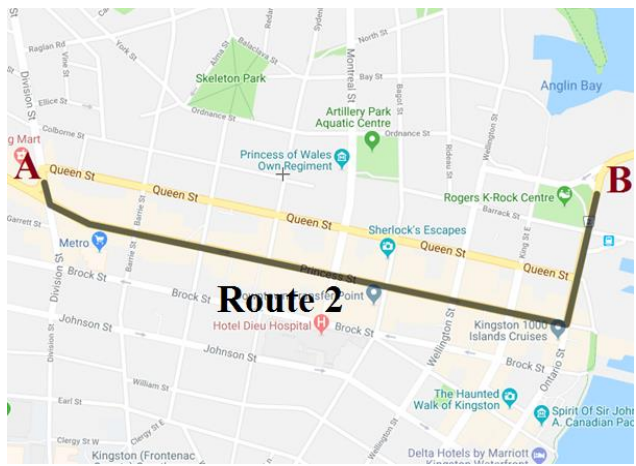

Fig8. Route suggestion provided by iDriveSense.

Comprehensively, the presented results show the significance of considering road information quality in dynamic route planning. As the requests for safety and comfort trips have introduced new metrics in route suggestions. Thus iDrivesense showed high capabilities in providing dynamic, safe and comfortable trips. However, roads quality are subject to change due to the effects of traffic and harsh weather. To sustain reliable dynamic route planning, continuous road segments assessments are enabled by iDrivesense.

IV. CONCLUSION

Future smart cities are required to consider numerous aspects to meet the expectations of their residents. Smooth and safe vehicle routing come on the top of the resident's demands due to their implications for their comfort and productivity on a daily bases. However, the popular route planning systems and service providers are not considering the road quality information while providing their trip planning services. In this paper, we presented "iDriveSense" a crowdsensing based system to enable such challenging demand. Our system benefits from the sensing capabilities of the vehicle motion sensors and the inertial sensors and GPS receivers to monitor road surface conditions. Accordingly, provides a cloud-based dynamic route planning services. The system was successfully able to operate independently or cooperatively with route planning service providers as Google Maps. The system can adequately asses the quality of road segments considering various aspects that affect the drivers' comfort and safety enabling efficient dynamic route planning while maintaining reasonable trip times and route distances.
.